\documentclass[12pt]{article}
\usepackage{wrapfig}
\usepackage{epsfig}
\usepackage{hyperref}
\usepackage{amssymb}
\usepackage{amsmath,mathtools}
\usepackage{amsfonts}
\usepackage{latexsym}
\usepackage{wasysym}
\usepackage{multirow}
\usepackage{fixmath}
\usepackage{color}
\usepackage{stackrel}
\usepackage{txfonts}                                                            \usepackage{centernot}
\usepackage{slashed}
\usepackage{bbm}
\usepackage[low-sup]{subdepth}
\usepackage{hyperref}

\newcommand{\bea}{\begin{eqnarray}}
\newcommand{\eea}{\end{eqnarray}}
\newcommand{\bite}{\begin{itemize}}
\newcommand{\eite}{\end{itemize}}

\newcommand{\MSbar}{\,\overline{\!\rm MS\!}\;}
\date{}

\begin{document}
\title{
\vspace{-1.5cm} 
\flushleft
\phantom{\normalsize DESY 24-160} \\
%\vspace{-0.35cm}
%{\normalsize Edinburgh 2017/04} \\
%\vspace{-0.35cm}
%{\normalsize Liverpool LTH 1122} \\
%\vspace{-0.35cm}
%{\normalsize March 2017} \\
\vspace{0.75cm}
\centering{\bf QCD Lambda Parameter from Gradient Flow}\\[0.5em]}

\author{Gerrit Schierholz\\[1.25em] Deutsches Elektronen-Synchrotron DESY,\\ Notkestr. 85, 22607 Hamburg, Germany\\
  and\\
  II. Institut f\"ur Theoretische Physik, Universit\"at Hamburg\\
Luruper Chaussee 149, 22761 Hamburg, Germany}

\maketitle
%\vspace*{-0.5cm}

\begin{abstract}
Starting from the running coupling in the gradient flow scheme, the QCD lambda parameter $\Lambda_{\,\overline{\!\rm MS\!}\;}$ is determined analytically with respect to the reference scale $w_0$ in the pure gauge theory. A key element is the gluon condensate, which, being invariant under RG transformations, establishes a formal relationship between $\Lambda_{\MSbar}$ and $w_0$. Good agreement with numerical estimates is found.
\end{abstract}
\vspace*{0.25cm}

%\newpage
\section{Introduction}

The parameter $\Lambda$ is one of the fundamental quantities of QCD. It sets the scale for confinement and the running coupling constant $\alpha(\mu)$. In the pure gauge theory $\Lambda$ is the only mass parameter. It is also the only free parameter that determines the entire spectrum from the ultraviolet to the infrared. A lattice calculation of $\Lambda$ requires the selection of a reference scale that can be computed accurately. Preferred scales are the gradient flow scales $w_0$ and $\sqrt{t_0}$. The lambda parameter is renormalization scheme dependent. It is common practice to express $\Lambda$ in the $\MSbar$ scheme, which requires precise knowledge of the appropriate conversion factor.

The gradient flow is the perfect setting for this task. It is a multi-scale problem involving the passage from short-distance scales to the strongly coupled regime at long distances marked by $\Lambda$. The framework for dealing with physical problems involving different energy scales is the renormalization group flow. Exact renormalization group transformations are very difficult to implement numerically. The gradient flow~\cite{Narayanan:2006rf,Luscher:2010iy} provides a powerful alternative for scale setting, with no need for costly ensemble matching. It can be regarded as a particular, infinitesimal realization of the coarse-graining step of momentum space renormalization group transformations~\cite{Luscher:2013vga,Makino:2018rys,Abe:2018zdc,Carosso:2018bmz} \`a la Wilson~\cite{Wilson:1973jj}, Polchinski~\cite{Polchinski:1983gv} and Wetterich~\cite{Berges:2000ew}, which leaves the long-distance physics unchanged. A particular example of an operator that is invariant under the gradient flow is the gluon condensate, which analytically connects $\Lambda$ with the root of the flow time, $\sqrt{t}$. 

In this Letter I will compute the lambda parameter in relation to $w_0$. The calculation is restricted to the pure SU(3) gauge theory so far. At the end of the Letter I will comment on the prospect of extending the computation to QCD with dynamical quarks.

\section{Gradient flow and running coupling}

The gradient flow~\cite{Narayanan:2006rf,Luscher:2010iy} evolves the gauge field along the gradient of the action. The flow of gauge fields is defined by
\begin{equation}
\partial_{\,t}\,B_\mu(t,x) = D_\nu G_{\mu\nu}(t,x) \,, \quad G_{\mu\nu} = \partial_\mu\,B_\nu -\partial_\nu\,B_\mu + [B_\mu, B_\nu] \,,
\end{equation}
where $D_\mu$ is the covariant derivative and $B_\mu(t=0,x) = A_\mu(x)$ is the initial gauge field generated by the action
\begin{equation}
  S_G = \frac{1}{2 g_0^2} \int d^4x\, \textrm{Tr}\, F_{\mu\nu} F_{\mu\nu} \,,  
\end{equation}
where $g_0$ is the bare coupling constant, which is kept fixed. The renormalization scale is set to $\mu=1/\sqrt{8t}$, where $\sqrt{8t}$ is the smoothing range over which $B_\mu$ is averaged. 
The expectation value of the energy density
\begin{equation}
  E = \frac{1}{4}\, G_{\mu\nu}^a\,  G_{\mu\nu}^a 
\end{equation}
with $G_{\mu\nu} = G_{\mu\nu}^a\, \tau^a/2$ defines a renormalized coupling in the gradient flow scheme~\cite{Luscher:2010iy,Harlander:2016vzb},
\begin{equation}
\alpha_{GF}(\mu) =  \frac{4 \pi}{3}\, t^2 \langle E(t)  \rangle \,,
\label{coupling}
\end{equation}
which analytically continues the beta function
\begin{equation}
  \frac{\partial\, \alpha_{GF}(\mu)}{\partial \ln \mu} = - 2 \,\frac{\partial\, \alpha_{GF}(\mu)}{\partial \ln t} = \beta_{GF}(\alpha_{GF}) 
  \label{beta}
\end{equation}
into the nonperturbative regime. Integrating (\ref{beta}) gives the familiar result
\begin{equation}
  \frac{\Lambda_{GF}}{\mu}=\left[4\pi b_0\alpha_{GF}(\mu)\right]^{-\frac{b_1}{2\,b_0^2}}\,\exp\left\{-\frac{1}{8\pi b_0\alpha_{GF}(\mu)}-\int_0^{\alpha_{GF}(\mu)} \!\! d\alpha  \,\left[ \frac{1}{\beta_{GF}(\alpha)}+\frac{1}{8\pi b_0\alpha^2} -\frac{b_1}{2b_0^2\alpha}\right]\right\} \,.
  \label{lambdamu2}
\end{equation}
Formally, $\Lambda_{GF}$ is a constant of integration. Actually, it sets the scale for the mass gap and the like. Equation~(\ref{lambdamu2}) simplifies to
\begin{equation}
 \frac{\Lambda_{GF}}{\mu}=\exp\left\{-\int^{\alpha_{GF}(\mu)}_{\alpha_{GF}(\Lambda_{GF})}  \,\frac{d\alpha}{\beta_{GF}(\alpha)}\right\}\,,
  \label{lambdamu}
\end{equation}
using 
\begin{equation}
  \frac{b_1}{2 b_0^2} \log{\left[4\pi b_0\alpha_{GF}(\Lambda_{GF})\right]} + \frac{1}{8\pi b_0\alpha_{GF}(\Lambda_{GF})} + \int_0^{\alpha_{GF}(\Lambda_{GF})} \!\! d\alpha  \,\left[ \frac{1}{\beta_{GF}(\alpha)}+\frac{1}{8\pi b_0\alpha^2} -\frac{b_1}{2b_0^2\alpha}\right] = 0 \,.
\end{equation}
It should be noted that $d\alpha/\beta(\alpha)$ is invariant under scheme transformations $GF \rightarrow S$. It follows that $\alpha_{GF}(\Lambda_{GF}) = \alpha_S(\Lambda_S)$. Under a change of renormalization scheme we then have
\begin{equation}
  \log {\frac{\Lambda_S}{\Lambda_{GF}}} = - \int_{\alpha_{GF}(\mu)}^{\alpha_{S}(\mu)} \, \frac{d\alpha}{\beta(\alpha)} \,.
\end{equation}
This must be true for all values of $\mu$. Taking $\mu$ to infinity, we obtain the one-loop result
\begin{equation}
  \Lambda_S = \exp\left\{\frac{t_1}{2b_0}\right\} \Lambda_{GF} \,, \quad \frac{1}{\alpha_S(\mu)} = \frac{1}{\alpha_{GF}(\mu)} - 4\pi \, t_1 \,.
  \label{con}
\end{equation}
The conversion (\ref{con}) is known for a variety of schemes $S$. For the $\MSbar$ scheme the result is~\cite{Luscher:2010iy}
\begin{equation}
  \frac{\Lambda_{\MSbar}}{\Lambda_{GF}} = 0.534 \,.
  \label{rat}
\end{equation}

\section{Gluon condensate}

Operators consisting of polynomials of $B_\mu(t,x)$, and its derivatives, are renormalized and finite for $t > 0$, subject to $t$ is large enough~\cite{Luscher:2011bx}. The running coupling and the energy density  combine to form the gluon condensate, which in the gradient flow scheme at finite flow time is defined by 
\begin{equation}
  G = \frac{\alpha_{GF}(\mu)}{\pi}\, \langle G_{\mu\nu}^a G_{\mu\nu}^a\rangle \,.
\end{equation}
The gluon condensate $G$ is scheme dependent, which is reflected in the fact that it is not directly related to any observable. Using the small flow-time expansion, $G$ lends itself to the definition of the renormalized gluon condensate at $t=0$~\cite{Artz:2019bpr}. This is then converted to the familiar condensate in the $\MSbar$ scheme~\cite{Shifman:1978bx}.
%The gluon condensate $G$ is a central element of QCD sum rules~\cite{Shifman:1978by}. It is also a favored object for studying the structure of the QCD vacuum.
Most importantly, $G$ is invariant under changes of the scale parameter $\mu$. This has been demonstrated both perturbatively~\cite{Kluberg-Stern:1974iel,Tarrach:1981bi} (the two-loop result~\cite{Tarrach:1981bi} is scale invariant, but not renormalization group invariant) and nonperturbatively~\cite{Shifman:1978bx,Novikov:1979va,Graziani:1984cs,Suzuki:2018vfs}. I provide a proof in~\cite{GS} that is derived directly from the gradient flow. In the gradient flow scheme this reads%~\cite{Harlander:2016vzb,Artz:2019bpr}
\begin{equation}
  t \, \frac{\partial}{\partial \,t} G = 0 \,.
\end{equation}
At flow time zero $G$ can be connected to the vacuum expectation value of the trace of the energy momentum tensor $\Theta_{\mu\nu}$, which sometimes is taken for the gluon condensate. The energy momentum tensor is obtained by taking the functional derivative of the action with respect to the metric tensor $g_{\mu\nu}$. However, no closed expression is known for finite flow time beyond perturbation theory~\cite{Suzuki:2013gza}.  

In actual lattice calculations $G$ is found to suffer from ultraviolet divergences due to mixing with the unit operator. In an unprecedented calculation the ultraviolet divergent contributions have been computed to $20th$~\cite{Horsley:2012ra} and $32nd$ order~\cite{Bali:2014sja} in numerical stochastic perturbation theory. Under the gradient flow the perturbative contributions (although finite) will gradually be eliminated as the flow time $t$ is increased, leaving behind the truly nonperturbative part. In the following we exclude the region near flow time zero. How large $t$ has to be for the perturbative contributions to have vanished needs to be seen.
Writing
\begin{equation}
  G = \frac{\alpha_{GF}(\mu)}{\pi}\, 4 \, \langle E(t) \rangle \,,
\end{equation}
it follows that 
\begin{equation}
    t \frac{\partial}{\partial \, t} \, \alpha_{GF}(\mu) \, \langle E(t) \rangle = - \frac{1}{2}\, \beta_{GF} \, \langle E(t) \rangle + \alpha_{GF}(\mu) \; t \frac{\partial}{\partial \, t} \,\langle E(t) \rangle = 0 \,.
    \label{rel2}
\end{equation}
Replacing $\langle E(t) \rangle$ with $3 \alpha_{GF}(\mu)/(4\pi t^2)$ in the right-hand expression, this leads to the beta function
\begin{equation}
  \beta_{GF} = - 2 \,\alpha_{GF} 
  \label{alphabeta}
\end{equation}
for corresponding values of $t$. The solution of (\ref{alphabeta}) has the form 
\begin{equation}
  \alpha_{GF}(\mu) = 8\, \Lambda^2 t = \frac{\Lambda^2}{\mu^2} \,. %\propto \frac{t}{a^2} = \frac{1}{8 \mu^2 a^2} \,.
    \label{ex}
\end{equation}
To capitalize on this, we will need to know the scale factor $\Lambda^2 = \alpha_{GF}(\mu)\, \mu^2$.

\begin{figure}[b!] \vspace*{-0.25cm}
  \begin{center}
  \epsfig{file=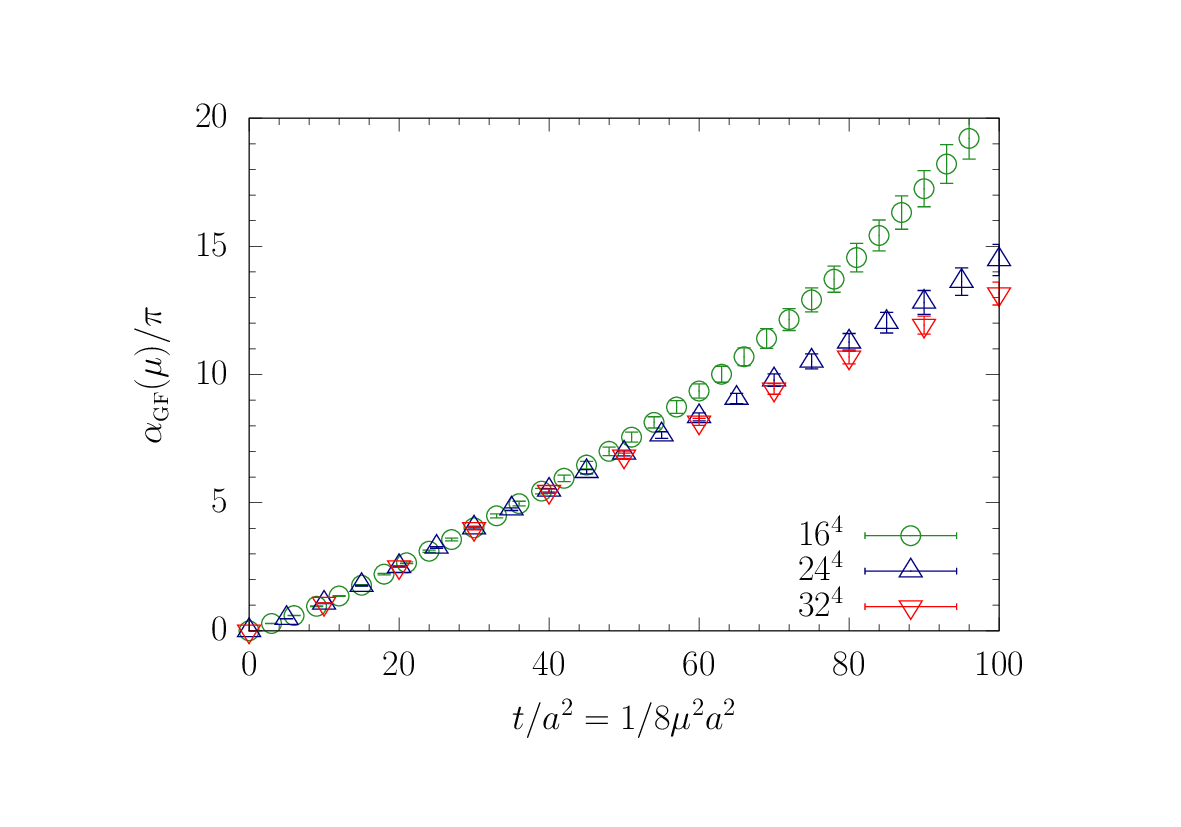,width=10cm,clip=}
  \end{center} \vspace*{-0.75cm}
  \caption{The running coupling $\alpha_{GF}(\mu)$ for the Wilson action at $\beta=6.0$ and lattice spacing $a=0.082\,\textrm{fm}$.}
  \label{fig1}
\end{figure}

In Fig.~\ref{fig1} I compare the prediction (\ref{ex}) with previous numerical results~\cite{Nakamura:2021meh}. The numbers refer to the Wilson action
\begin{equation}
  S = \beta \sum_{x,\,\mu < \nu} \Big( 1 - \frac{1}{3}\, \mathrm{Re}\, \mathrm{Tr}\; U_{\mu\nu}(x)\Big)
\end{equation}
at $\beta = 6.0$, corresponding to a lattice spacing $a = 0.082\, \textrm{fm}$, and the clover definition of $G_{\mu\nu}$. To test for finite volume effects, the calculations were done on three lattice volumes, $16^4$, $24^4$ and $32^4$, with $5,000$ configurations each. Finite volume effects are expected to be small for $\sqrt{8t} \ll L$, where $L$ is the linear extent of the lattice. On the larger lattices the data are compatible with a linear rise in $t$, as expected. On the smaller lattice finite volume corrections become clearly visible at $\sqrt{8t} \gtrsim L$. At small flow times the coupling is known~\cite{Cheng:2014jba,Fodor:2014cpa} to be sensitive to $O(a^2)$ corrections. Added to this, as the flow time increases, $\langle E\rangle$ splits into discrete values depending on the topological charge $Q$. Asymptotically $\langle E\rangle \propto |Q|$~\cite{Nakamura:2021meh}, so that the higher charges become increasingly important. This calls for a precise knowledge of $\langle E\rangle$ at large $|Q|$, which requires ultra-high statistics and sufficiently smooth gauge fields to begin with.

\begin{figure}[t!] \vspace*{-0.75cm}
  \begin{center}
  \epsfig{file=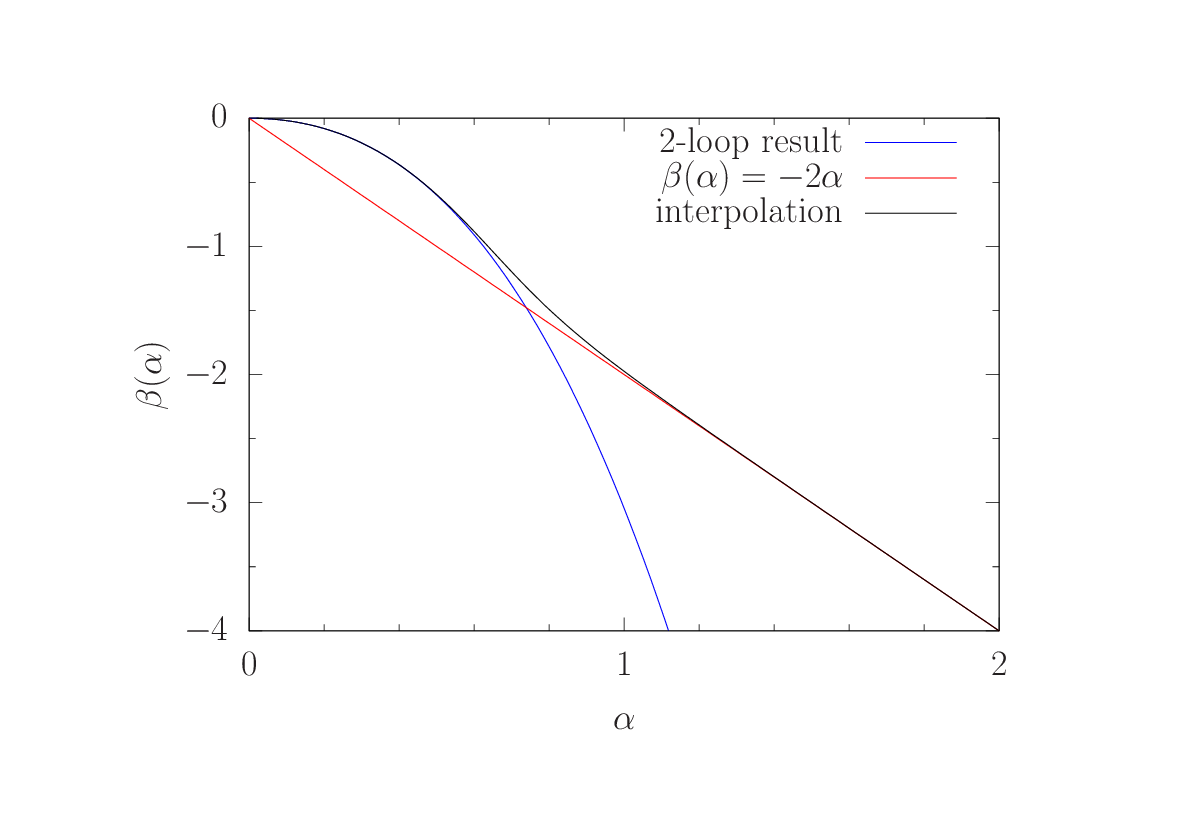,width=11.75cm,clip=}
  \end{center} \vspace*{-1.5cm}
  \caption{The beta function that interpolates between the two-loop result $\beta(\alpha)=-8\pi b_0 \alpha^2 - 32\pi^2 b_1 \alpha^3$, which applies to small values of $\alpha$, and the `asymptotic' result $\beta(\alpha)=-2\alpha$, which applies to large values of $\alpha$. The continuous interpolating line corresponds to $n=3$.}
  \label{fig2}
\end{figure}

\section{Scale setting and $\Lambda$ parameter}

\begin{figure}[t!] \vspace*{-0.5cm}
  \begin{center}
  \epsfig{file=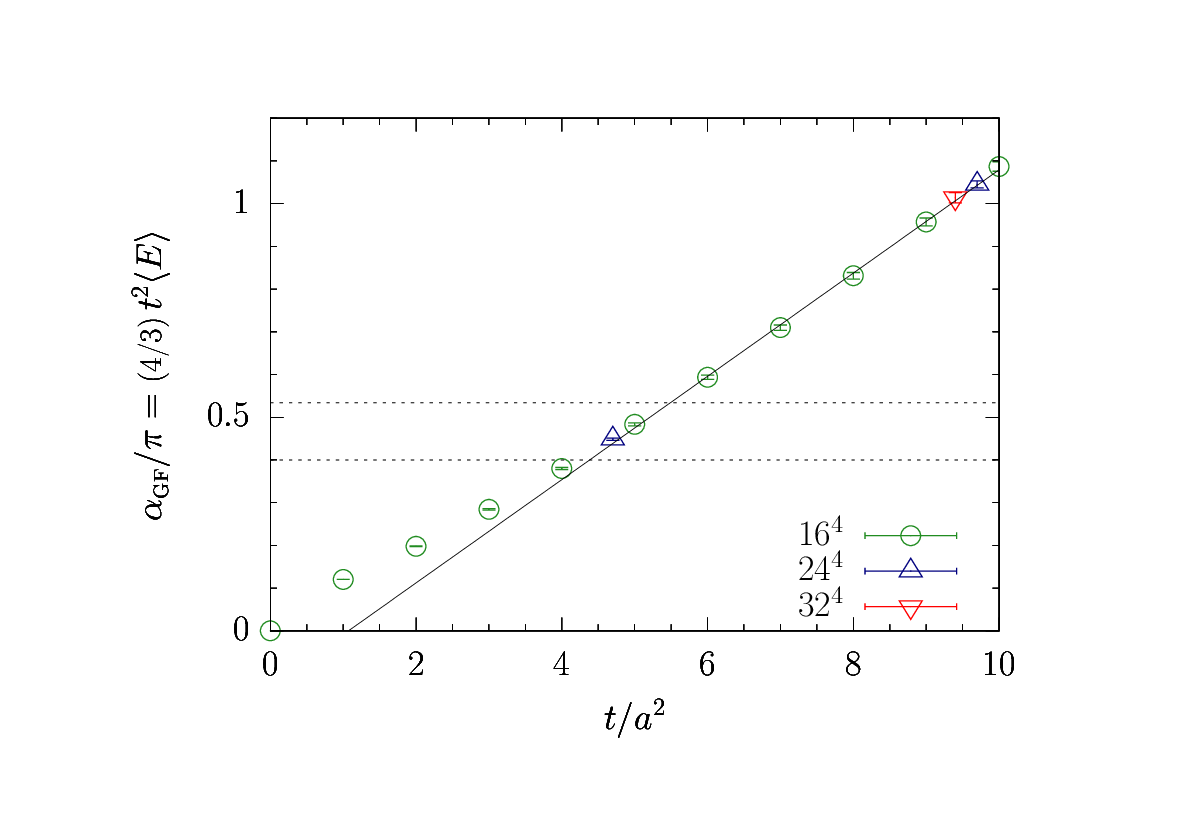,width=10cm,clip=}
  \end{center} \vspace*{-0.75cm}
  \caption{The running coupling $\alpha_{GF}(\mu) = (4/3)\, t^2\langle E\rangle$ for small values of $t/a^2$. The dashed lines represent $c = 0.3$ (bottom) and $0.4$ (top), respectively. The solid line is a fit to the lattice data~\cite{Nakamura:2021meh} for $t/a^2 \geq 5$. From the line we read off $w_0(0.3)/a = 1.818$ and $a \Lambda_{GF} \approx 0.218$, which gives, per definition, $w_0(0.3) \Lambda_{GF} = 0.396$, as predicted by (\ref{w0lam}).}
  \label{fig3}
\end{figure}

While for large values of $\alpha_{GF}$ the beta function takes the form $\beta_{GF} = -2 \alpha_{GF} \equiv \beta_{GF}^{\,>}$, for small values it can be approximated by the two-loop expression $\beta_{GF} = -4\pi b_0 \alpha_{GF}^2 - 32\pi^2 b_1 \alpha_{GF}^3 \equiv \beta_{GF}^{\,<}$. This is sketched in Fig.~\ref{fig2}. A smooth interpolation that connects the two beta functions is given, for example, by
\begin{equation}
  \frac{1}{\beta_{GF}}=-\left[\,\left(\frac{1}{\beta_{GF}^{\,<}}\right)^{2n} + \left(\frac{1}{\beta_{GF}^{\,>}}\right)^{2n}\,\right]^{\,\raisebox{-3pt}{$\scriptstyle 1/2n$}} \,,
\end{equation}
which is represented by the connecting line. At $\alpha_{GF}=0$ the integral (\ref{lambdamu2}) over the two-loop beta function vanishes,
\begin{equation}
  \int_0 d\alpha  \,\left[ \frac{1}{\beta_{GF}(\alpha)}+\frac{1}{8\pi b_0\alpha^2} -\frac{b_1}{2b_0^2\alpha}\right] = 0 \,,
  \label{zero}
\end{equation}
so that
\begin{equation}
   \log {\frac{\Lambda_{GF}}{\mu}} = - \frac{b_1}{2 b_0^2} \log{\left[4\pi b_0\alpha_{GF}(\mu)\right]} - \frac{1}{8\pi b_0\alpha_{GF}(\mu)} - \int^{\alpha_{GF}(\mu)} \!\!\! d\alpha  \,\left[ \frac{1}{\beta_{GF}(\alpha)}+\frac{1}{8\pi b_0\alpha^2} -\frac{b_1}{2b_0^2\alpha}\right] \,.
\end{equation}
For larger values of $\alpha_{GF}$ this must match the `asymptotic' result
\begin{equation}
    - \int^{\alpha_{GF}(\mu)} \!\! d\alpha  \, \frac{1}{\beta_{GF}(\alpha)} = \frac{1}{2} \log {\alpha_{GF}(\mu)}\,.
\end{equation}
A suitable matching point is $\mu = \Lambda_{GF}$, for example. (See Fig.~\ref{fig3}.) From this follows that $\Lambda = \Lambda_{GF}$ and 
\begin{equation}
  \alpha_{GF}(\mu) = \frac{\Lambda_{GF}^2}{\mu^2} \,.
  \label{ex2}
\end{equation}
This offers the possibility of determining $\Lambda_{GF}$ directly from the gradient flow, which is the key observation of this note.

It is practice to express dimensionful results from the lattice in units of a common scale parameter. Two such parameters are derived from the gradient flow, $\sqrt{t_0}$ and $w_0$, which can be determined precisely with moderate effort. Setting
\begin{equation}
  F(t) = t^2\, \langle E(t)\rangle \,,
\end{equation}
the $\sqrt{t_0}$ scale is defined by
\begin{equation}
  F(t)\,|_{t=t_0(c)} = c \,.
\end{equation}
Alternatively, the $w_0$ scale is defined by
\begin{equation}
  t \frac{\partial}{\partial t} F(t)\,|_{t=w_0^2(c)} = c \,.
  \label{w0}
\end{equation}
In both definitions $c$ is a constant, conventionally taken as $c = 0.3$. We require a value of $c$ such that  $a \ll \sqrt{t_0}$ to evade $O(a^2)$ effects and issues of renormalization~\cite{Luscher:2011bx}, and $\sqrt{8t_0} \ll L$ to exclude finite volume corrections.

For flow times in the linear range we can write
\begin{equation}
  F(t) = \Delta_0 + \Delta \, t \,,
\end{equation}
where $\Delta_0$ is an unknown constant that can be either negative as in our case or positive~\cite{Asakawa:2015vta}. We then have
\begin{equation}
  \begin{split}
    \Delta_0 + \Delta \, t_0(c) &= c \,, \\
    \Delta w_0^2(c) &= c \,.
  \end{split}
\end{equation}
As a result
\begin{equation}
   t_0(c) = w_0^2(c) - \frac{\Delta_0}{\Delta} \,.
\end{equation}
While $w_0^2(c)/c$ is independent of $c$ in the linear regime, and a true physical scale (proportional to $1/\Lambda^2$), $t_0(c)/c$ is generally not. Only for large values of $c$ can we expect that $t_0(c) \simeq w_0^2(c)$. 

In Fig.~\ref{fig3} I show a detailed view of Fig.~\ref{fig1} for smaller values of $t/a^2$. At this lattice spacing, $a = 0.082\,\textrm{fm}$, the nonperturbative linear regime begins approximately at $F(t) = 0.3$, rendering the value $c = 0.3$ just acceptable. It shows that $c = 0.4$ or larger would be better in this case. The situation improves for smaller lattice spacings, with $c = 0.3$ moving into the linear regime~\cite{Asakawa:2015vta}. We suggest to fit the (approximately?) linear part of $F(t)$ by a straight line as shown in Fig.~\ref{fig3}. This gives
\begin{equation}
  \frac{\partial F(t)}{\partial t} = \frac{c}{w_0^2(c)} \,,
\end{equation}
which eliminates the need for $c$. In contrast, $c/t_0(c) = \partial F(t)/\partial t + \Delta_0/t_0(c)$ becomes independent of $c$ only for $t_0(c) \gg |\Delta_0|$.

Given the analytic expression (\ref{ex2}) of $\alpha_{GF}(\mu)$, we have
\begin{equation}
  \frac{\partial\, F}{\partial \, t} = \frac{3}{4\pi}\, \frac{\partial\, \alpha_{GF}}{\partial \, t} = - \frac{3}{\pi}\, \mu^3\, \frac{\partial\, \alpha_{GF}}{\partial \, \mu}\ = \frac{6}{\pi} \, \Lambda_{GF}^2\,.
\end{equation}
Combining this result with the definition of $w_0$ in (\ref{w0}), we obtain
\begin{equation}
  w_0(c) \,\Lambda_{GF} = \sqrt{\frac{c \,\pi}{6}} \,.
  \label{w0lam}
\end{equation}
The preferred scheme in phenomenological applications is  $\MSbar$. Using the relation (\ref{rat}), we finally obtain
\begin{equation}
  w_0(c) \,\Lambda_{\MSbar} = 0.534 \, \sqrt{\frac{c \,\pi}{6}} \,.
\end{equation}
For $c = 0.3$ the result is
\begin{equation}
  w_0 \,\Lambda_{\MSbar} = 0.212 \,.
  \label{final03}
\end{equation}
This number is in good agreement with recent results from the literature:
\begin{equation}
  \begin{split}
    & \textrm{\!Ref.}\\
    w_0 \Lambda_{\MSbar} = 0.225(4) \quad\quad & \cite{Asakawa:2015vta}\\[0.5em]
    w_0 \Lambda_{\MSbar} = 0.215(2) \quad\quad & \cite{Kitazawa:2016dsl}\\[0.5em]
    \sqrt{t_0} \Lambda_{\MSbar} = 0.220(4) \quad\quad & \cite{DallaBrida:2019wur}\\[0.5em]
%    w_0 \Lambda_{\MSbar} = 0.217(3) \quad\quad & \cite{Nakamura:2021meh}\\[0.5em]
    \sqrt{t_0} \Lambda_{\MSbar} = 0.219(4) \quad\quad & \cite{Hasenfratz:2023bok}\\
  \end{split}
  \label{ref}
\end{equation}
This applies to the last two entries as well. The factor to convert $\sqrt{t_0}$ to $w_0$ was found to be $w_0/\sqrt{t_0} = 1.02(2)$ for $c \geq 0.3$ and a wide range of lattice spacings $w_0(0.3)/a \gtrsim 2.9$~\cite{Asakawa:2015vta}. For comparison, at our lattice spacing $w_0(0.3)/a \approx 2.1$. In (\ref{ref}) only the scale setting parameters were computed using the gradient flow, whereas the lambda parameters were determined by matching the tadpole improved bare coupling to the $\MSbar$ scheme and finite size scaling techniques, which explains the deviations from (\ref{final03}). The deviations from each other are within the ballpark of $O(a^2)$ and finite size effects. 

\section{Conclusions}

The determination of the QCD lambda parameter has a long history. Unlike previous calculations, our work is based on the infrared behavior of the running coupling, which is most sensitive to lambda and provides a direct link to the nonperturbative quantities of the theory, such as the string tension~\cite{Nakamura:2021meh}. The crucial point here is that the running coupling is a linearly rising function of flow time $t$ for $t \gg a^2$, which is a direct consequence of the gluon condensate being scale invariant and is supported by numerical simulations. The most important result is that, to determine the lambda parameter, we only require knowledge of the scale setting parameter $w_0$. We have demonstrated how $w_0$ should be calculated correctly in practice.

All this is under the premise that the gradient flow is a realization of momentum space renormalization group transformations. Significant work has been done to establish the connection between the gradient flow and renormalization group transformations~\cite{Luscher:2013vga,Makino:2018rys,Abe:2018zdc,Carosso:2018bmz}. %An independent evaluation of the energy momentum tensor under the gradient flow~\cite{Suzuki:2013gza} for large flow times would be a valuable addition to this work.

As a byproduct we obtain the (renormalized) gluon condensate $G$. The result is $G = 19.45\, \Lambda_{\MSbar}^4$. This is broadly consistent with recent results from the plaquette, where the perturbative contribution has been subtracted to an unprecedented order~\cite{Horsley:2012ra,Bali:2014sja}. A formal derivation will be given elsewhere. The linear increase of the running coupling beyond bounds is sometimes referred to as infrared slavery, the flip side of asymptotic freedom. Our result that confinement simply follows from a nonvanishing gluon condensate is a valuable addition to the modelling of the QCD vacuum.

While the result for the pure gauge theory is interesting on its own, the final goal is to extend the calculations to multiple quark flavors. One way to achieve this is by using the `nonperturbative decoupling method'~\cite{DallaBrida:2022eua}, starting from the now exact, zero-flavor result (\ref{final03}). A direct calculation of $\Lambda_{\MSbar}$ (say) for three light flavors should also be possible. In this case the task is to align $\Lambda_{\MSbar}$ with the gradient flow scale $w_0$ in the corresponding limit.

\end{document}